\documentclass{appolb}
\title{AN ANALYTICAL STRONG COUPLING APPROACH\\ IN DYNAMICAL MEAN-FIELD THEORY%
\thanks{Presented at the XII School of Modern Physics on Phase Transitions and
Critical Phenomena, L\c{a}dek Zdr\'oj, Poland, June 21--24, 2001}}

\author{A.M. Shvaika
\address{Institute for Condensed Matter Physics,
Natl. Acad. Sci. Ukr. \\
1~Svientsitskii Str., UA-79011 Lviv, Ukraine}}

\date{June 21, 2001}

\headauthor{A.M. Shvaika}
\headtitle{An Analytical Strong Coupling Approach \dots}
\Redakcjatrue

\begin{document}

\setcounter{page}{3415}

\maketitle

\begin{abstract}
In the limit of infinite spatial dimensions a thermodynamically
consistent theory of the strongly correlated electron systems,
which is valid for arbitrary value of the Coulombic interaction
($U<\infty$), is built. For the Hubbard model the total auxiliary
single-site problem exactly splits into four subspaces, which
describe Fermi and non-Fermi liquid components. Such analytical
approach allows to construct different thermodynamically
consistent approximations: alloy-analogy approximation,
Hartree-Fock approximation and further, that describes the
self-consistent renormalization of the bosonic excitations
(magnons and doublons).
\end{abstract}


\vfill

In the last decade a lot of the rigorous results in the theory of
strongly correlated electron systems are connected with the
development of the Dynamical Mean-Field Theory (DMFT) proposed by
Metzner and Vollhardt \cite{MetznerVollhardt} for the Hubbard
model (see also \cite{DMFTreview} and references therein). It maps
lattice problem on the effective single impurity Anderson model
with the generalized partition function
\begin{equation}\label{eq1}
  \hat{\rho} = e^{-\beta \hat{H}_{0}} T \!\exp
  \left\{\!-\!\int\nolimits_{0}^{\beta} \!\!d\tau
  \!\int\nolimits_{0}^{\beta}\!\!d\tau' \sum\limits_{\sigma}
  \zeta_{\sigma}(\tau-\tau') a_{\sigma}^{\dag}(\tau) a_{\sigma}(\tau')
  \right\},
\end{equation}
where $\zeta_{\sigma}(\tau-\tau')$ is an auxiliary Kadanoff-Baym
field (single-site hopping) and for the Hubbard model
\begin{equation}\label{eq2}
  \hat{H}_0 = Un_{\uparrow} n_{\downarrow} - \mu(n_{\uparrow} +
  n_{\downarrow}) - h(n_{\uparrow} - n_{\downarrow}) =
  Un_{\uparrow} n_{\downarrow} - \sum_{\sigma} \mu_{\sigma} n_{\sigma}
  = \sum_p \lambda_p \hat{X}^{pp},
\end{equation}
and there are no restrictions on the $U$ value within this theory.
Moreover, some classes of binary-alloy-type models (\emph{e.g.}, the
Falicov-Kimball model) can be studied exactly within DMFT
\cite{BrandtMielsch}. But in the case of the Hubbard model, the
treatment of the effective single impurity Anderson model is very
complicated and mainly computer simulations are used, which calls
for the development of the analytical approaches \cite{Gebhard}.
For the partition function (\ref{eq1}) such approach can be build
within a perturbation theory expansion in terms of the electron
hopping using a diagrammatic technique for Hubbard operators
\cite{ShvaikaPRB}, which is based on the corresponding Wick
theorem \cite{Slobodyan}. In present article we supplement such
strong-coupling approach by the consideration of the bosonic
fluctuations.

Consecutive pairing of all off-diagonal Hubbard operators $X^{pq}$
is performed until we get the product of the diagonal operators
only. For the single impurity problem (\ref{eq1}) and (\ref{eq2})
all diagonal operators act at the same site, their products can be
reduced to the single operators and for the Hubbard model our
problem exactly splits into four subspaces with ``vacuum states''
$|p\rangle=|0\rangle$, $|2\rangle$, $|\uparrow\rangle$ and
$|\downarrow\rangle$ and only excitations, as fermionic, as
bosonic, around these ``vacuum states'' are allowed. Finally, for
the grand canonical potential we get \cite{ShvaikaPRB}
\begin{equation}\label{eq3}
  \Omega_{a} = -\frac{1}{\beta} \ln \sum_{p} e^{-\beta\Omega_{(p)}},
\end{equation}
where $\Omega_{(p)}$ are the ``grand canonical potentials'' for
the subspaces. Now we can find the single-electron Green functions
by
\begin{equation}\label{eq4}
  G_{\sigma}^{(a)} (\tau-\tau') =
  \frac{\delta\Omega_{a}}{\delta \zeta_{\sigma}(\tau-\tau')} =
  \sum_{p} w_{p} G_{\sigma(p)} (\tau-\tau'),
\end{equation}
where $G_{\sigma(p)} (\tau-\tau')$ are single-electron Green
functions for the subspaces characterized by the ``statistical
weights''
\begin{equation}\label{eq5}
  w_{p} =\frac
  {e^{-\beta\Omega_{(p)}}}{\sum_{q}e^{-\beta\Omega_{(q)}}}.
\end{equation}
We can introduce irreducible parts of Green's functions in
subspaces $\Xi_{\sigma(p)}$ by
\begin{equation}\label{eq6}
  G_{\sigma(p)}^{-1} (\omega_{\nu}) =
  \Xi_{\sigma(p)}^{-1}(\omega_{\nu}) - \zeta_{\sigma}(\omega_{\nu})
\end{equation}
and self-energies in subspaces $\Sigma_{\sigma(p)}$ (Dyson
equation for the irreducible parts)
\begin{equation}\label{eq7}
  \Xi_{\sigma(p)}^{-1}(\omega_{\nu})=i\omega_{\nu}+\mu_{\sigma}-Un^{(0)}_{\bar\sigma(p)}
  -\Sigma_{\sigma(p)}(\omega_{\nu}),
\end{equation}
where $n^{(0)}_{\sigma(p)}=-\frac{d\lambda_p}{d\mu_{\sigma}}=0$
for $p=0,\bar\sigma$ and $1$ for $p=2,\sigma$. Here, self-energy
$\Sigma_{\sigma(p)}(\omega_{\nu})$ depends on the hopping integral
$\zeta_{\sigma'}(\omega_{\nu'})$ only through quantities
\begin{equation}\label{eq8}
  \Psi_{\sigma'(p)} (\omega_{\nu'}) = G_{\sigma'(p)} (\omega_{\nu'}) -
  \Xi _{\sigma'(p)} (\omega_{\nu'}).
\end{equation}

Now, one can reconstruct expressions for the grand canonical
potentials $\Omega_{(p)}$ in subspaces from the known structure of
Green functions:
\begin{eqnarray}\label{eq10}
  \Omega_{(p)} = \lambda_{p} & - & \frac{1}{\beta} \sum_{\nu\sigma}
  \ln\left[1 - \zeta_{\sigma}(\omega_{\nu})\Xi_{\sigma(p)}(\omega_{\nu})\right]
  \\ \nonumber
  & - & \frac{1}{\beta} \sum_{\nu\sigma} \Sigma_{\sigma(p)}(\omega_{\nu})
  \Psi_{\sigma(p)}(\omega_{\nu}) + \Phi_{(p)},
\end{eqnarray}
where $\Phi_{(p)}$ is some functional, such that its functional
derivative with respect to $\Psi$ produces the self-energy:
\begin{equation}\label{eq10p}
\beta\frac{\delta\Phi_{(p)}}{\delta\Psi_{\sigma(p)}(\omega_{\nu})}
=\Sigma_{\sigma(p)}(\omega_{\nu}).
\end{equation}
The second term in (\ref{eq10}) corresponds to the sum of the
fermionic single loop contributions whereas the next ones describe
different scattering processes.

From the grand canonical potential (\ref{eq3}) and (\ref{eq10}) we
get for mean values
\begin{eqnarray}\label{eq11}
  n_{\sigma}&=&-\frac{d\Omega_a}{d\mu_{\sigma}}=
  \sum_p w_p n_{\sigma(p)},
  \\ \nonumber
  n_{\sigma(p)}&=&n^{(0)}_{\sigma(p)}+
  \frac1{\beta}\sum_{\nu}\Psi_{\sigma(p)}(\omega_{\nu})
  -\frac{\partial\Phi_{(p)}}{\partial\mu_{\sigma}},
\end{eqnarray}
where in the last term the partial derivative is taken over
$\mu_{\sigma}$ not in quantities $\Psi$ (\ref{eq8}).

\emph{Falicov-Kimball model} correspond to the case of
$\zeta_{\downarrow}(\omega_{\nu}) \equiv 0$ and, as a result,
$\Phi_{(p)} \equiv 0$ and $\Sigma_{\uparrow(p)}(\omega_{\nu})
\equiv 0$ which immediately gives results of \cite{BrandtMielsch}
(see also \cite{ShvaikaJPS}).

Presented above equations allows one to construct different
thermodynamically consistent approximations.

The simplest \emph{alloy-analogy approximation}, which is a
zero-order approximation within the considered approach, is to put
$\Phi_{(p)} = 0$, $\Sigma_{\sigma(p)}(\omega_{\nu})=0$ and for the
Green's function for the single impurity problem one can obtain a
two-pole expression ($\lambda_{pq}=\lambda_{p}-\lambda_{q}$)
\begin{equation}\label{eq15}
  G_{\sigma}^{(a)}(\omega_{\nu})=\frac{w_0+w_{\sigma}}
            {i\omega_{\nu}-\lambda_{\sigma0}-\zeta_{\sigma}(\omega_{\nu})}+
            \frac{w_2+w_{\bar\sigma}}
            {i\omega_{\nu}-\lambda_{2\bar\sigma}-\zeta_{\sigma}(\omega_{\nu})}.
\end{equation}

\emph{Strong coupling Hartree-Fock approximation} takes into
account the first corrections into the self energy in the form
\begin{eqnarray}\label{eq17}
  \Sigma_{\sigma(p)}(\omega_{\nu})&=&U\left(n_{\sigma(p)}-n^{(0)}_{\sigma(p)}\right),
  \\ \nonumber
  n_{\sigma(p)} &=& n^{(0)}_{\sigma(p)} + \frac{1}{\beta}
  \sum\limits_{\nu} \Psi_{\sigma(p)}(\omega_{\nu}) ,
\end{eqnarray}
which gives for the Green's function a four-pole expression:
\begin{equation}\label{eq19}
  G^{(a)}_{\sigma}(\omega_{\nu})=\sum_p
  \frac{w_p}{i\omega_{\nu}+\mu_{\sigma}-Un_{\bar\sigma(p)}
  -\zeta_{\sigma}(\omega_{\nu})}.
\end{equation}
Now, grand canonical potentials in the subspaces (\ref{eq10}) are
determined by functional
\begin{equation}\label{eq20}
  \Phi_{(p)}=U \left(n_{\sigma(p)}-n_{\sigma(p)}^{(0)}\right)
  \left(n_{\bar\sigma(p)}-n_{\bar\sigma(p)}^{(0)}\right).
\end{equation}
Expression (\ref{eq19}) possesses the correct Hartree-Fock limit
for the small Co\-u\-lo\-m\-bic interaction $U$ and electron or
hole concentrations. On the other hand, in the same way as an
alloy-analogy solution, it describes the metal-insulator
transition with the change of $U$. In \cite{ShvaikaPRB,ShvaikaCMP}
it was shown that the main contributions come from the subspaces
$p=0$ or $2$, that describe the Fermi-liquid component, for the
low electron ($n\ll 1$) or hole ($2-n\ll 1$) concentrations,
respectively, (``overdoped regime'' of high-$T_c$'s) and
$p=\uparrow,\downarrow$, that describe the non-Fermi-liquid
component, close to half filling ($n\sim 1$) (``underdoped
regime''). At low temperatures the Fermi-liquid component is in
the ferromagnetic (Stoner) state, while the non-Fermi-liquid one
is antiferromagnetic close to half-filling.

In order to go \emph{beyond the Hartree-Fock approximation} one
have to consider, besides the fermionic excitations, also the
bosonic ones \cite{ShvaikaPRB} which correspond to the creation
and annihilation of the doublons (pairs of electrons), described
by the $\hat{X}^{20}$ and $\hat{X}^{02}$ operators, for subspaces
$p=0,2$ and magnons, described by the
$\hat{X}^{\uparrow\downarrow}$ and $\hat{X}^{\downarrow\uparrow}$
operators, for $p=\uparrow,\downarrow$. The single loop
contributions of such bosonic excitations can be summed up and one
can obtain
\begin{equation}\label{eq21}
  \Phi_{(p)}=\frac1{\beta}\sum_m \ln
  \left[1-U\left(1\pm U\tilde{D}_{\sigma\bar\sigma(p)}(\omega_m)\right)
  \tilde\chi_{\sigma\bar\sigma(p)}(\omega_m)\right] ,
\end{equation}
where
\begin{eqnarray}\label{eq22}
  \tilde{D}_{\sigma\bar\sigma(p)}(\omega_m)&=&\frac1{i\omega_m-\lambda_{20}-
  U\frac1{\beta}\sum_{\nu}
  \left[\Psi_{\sigma(p)}(\omega_{\nu})+\Psi_{\bar\sigma(p)}(\omega_{\nu})\right]},
  \\ \label{eq23}
  \tilde\chi_{\sigma\bar\sigma(p)}(\omega_m)&=&-\frac1{\beta}\sum_{\nu}
  \Psi_{\sigma(p)}(\omega_{\nu}) \Psi_{\bar\sigma(p)}(\omega_{m-\nu})
\end{eqnarray}
for subspaces $p=0,2$ and
\begin{eqnarray}\label{eq24}
  \tilde{D}_{\sigma\bar\sigma(p)}(\omega_m)&=&\frac1{i\omega_m-\lambda_{\sigma\bar\sigma}-
  U\frac1{\beta}\sum_{\nu}
  \left[\Psi_{\bar\sigma(p)}(\omega_{\nu})-\Psi_{\sigma(p)}(\omega_{\nu})\right]},
  \\ \label{eq25}
  \tilde\chi_{\sigma\bar\sigma(p)}(\omega_m)&=&-\frac1{\beta}\sum_{\nu}
  \Psi_{\sigma(p)}(\omega_{\nu}) \Psi_{\bar\sigma(p)}(\omega_{\nu-m})
\end{eqnarray}
for subspaces $p=\sigma,\bar\sigma$, which gives for mean values
the following expression
\begin{eqnarray}\label{eq26}
  n_{\sigma(p)}&=&n_{\sigma(p)}^{(0)}+\frac1{\beta}\sum_{\nu}\Psi_{\sigma(p)}(\omega_{\nu})
  \\ \nonumber
  &&\mp \frac1{\beta}\sum_m\frac{U^2\tilde{D}_{\sigma\bar\sigma(p)}^2(\omega_m)
  \tilde\chi_{\sigma\bar\sigma(p)}(\omega_m)}
  {1-U\left(1\pm U\tilde{D}_{\sigma\bar\sigma(p)}(\omega_m)\right)
  \tilde\chi_{\sigma\bar\sigma(p)}(\omega_m)}.
\end{eqnarray}
Now self-energy contains the frequency dependent part
\begin{equation}\label{eq27}
  \Sigma_{\sigma(p)}(\omega_{\nu})=U\left(n_{\bar\sigma(p)}-n_{\bar\sigma(p)}^{(0)}\right)+
  \tilde\Sigma_{\sigma(p)}(\omega_{\nu}),
\end{equation}
\begin{eqnarray}\nonumber
  \tilde\Sigma_{\sigma(p)}(\omega_{\nu})&=&U^2
  \frac1{\beta}\sum_m\frac{\left(1\pm U\tilde{D}_{\sigma\bar\sigma(p)}(\omega_m)\right)
  \tilde\chi_{\sigma\bar\sigma(p)}(\omega_m)\pm\tilde{D}_{\sigma\bar\sigma(p)}(\omega_m)}
  {1-U\left(1\pm U\tilde{D}_{\sigma\bar\sigma(p)}(\omega_m)\right)
  \tilde\chi_{\sigma\bar\sigma(p)}(\omega_m)}
  \\ \label{eq28}
  &&\times\left\{
  \begin{array}{rl}
  \Psi_{\bar\sigma(p)}(\omega_{m-\nu}), & \mbox{for } p=0,2 \\
  \Psi_{\bar\sigma(p)}(\omega_{\nu-m}), & \mbox{for } p=\sigma,\bar\sigma
  \end{array}
  \right. ,
\end{eqnarray}
that describes the contributions from the doublons for the Fermi
liquid com\-po\-nent ($p=0,2$) and magnons for the non-Fermi liquid
one ($p=\uparrow,\downarrow$) with the renormalized spectrum
determined by the zeros of denominator in (\ref{eq28}).


Expression (\ref{eq21}) for functional $\Phi_{(p)}$ has the same
form as the correction to free energy in the theory of the
Self-Consistent Renormalization (SCR) of spin fluctuations by
Moriya \cite{MoriyaBook}. But in our case it describes
contributions from the single-site bosonic (magnon or doublon)
fluctuations with specific renormalization functions different for
different subspaces. Spin fluctuations give the main contribution
close to half filling in the non-Fermi liquid regime but for small
electron ($n\ll1$) or hole ($2-n\ll1$) concentrations the
contributions from the doublon (charge) fluctuations must be taken
into account.

\end{document}